\documentclass[journal, final, twocolumn, 11pt]{IEEEtran}
\usepackage{cite}
\usepackage[pdftex]{graphicx}
\usepackage{amsmath, amsfonts, amssymb, amsthm}
\usepackage{mathtools}
\usepackage{cancel}
\usepackage{booktabs, threeparttable, multirow}
\usepackage{float}
\usepackage[noend]{algorithm2e}
\usepackage[caption=false, font=footnotesize]{subfig}
\usepackage{microtype}
\usepackage{anyfontsize}
\usepackage{cleveref}
\usepackage{xfrac}

\makeatletter
\patchcmd{\@algocf@start}
    {-1.5em}
    {0pt}
    {}{}
\makeatother

\crefformat{subfigure}{#2\onlyletter{#1}#3}
\crefrangeformat{subfigure}{(#3\onlyletter{#1}#4--#5\onlyletter{#2}#6)}
\crefmultiformat{subfigure}
  {(#2\onlyletter{#1}#3}
  {,~#2\onlyletter{#1}#3)}
  {,~#2\onlyletter{#1}#3)}
  {,~#2\onlyletter{#1}#3)}

\ExplSyntaxOn
\NewDocumentCommand{\onlyletter}{m}
 {
  \tl_set:Nx \l_tmpa_tl { #1 }
  \tl_item:Nn \l_tmpa_tl { -1 }
 }
\ExplSyntaxOff

\crefname{figure}{}{}

\newtheorem{theorem}{Theorem}[section]

\newtheorem{proposition}[theorem]{Proposition}
\newtheorem{definition}{Definition}[section]

\begin{document}
\bstctlcite{BSTcontrol}
\title{A Joint Energy and Differentially-Private Smart Meter Data Market}
\author{Saurab~Chhachhi,~\IEEEmembership{Member,~IEEE,
} and Fei~Teng,~\IEEEmembership{Senior Member,~IEEE,}%
\thanks{Manuscript received \today; revised \today. This work was supported by ESRC through LISS DTP (ES/P000703/1:2113082).
(Corresponding author Saurab Chhachhi)}
\thanks{Saurab Chhachhi and Fei Teng are with the Control and Power Group, Department of Electrical and Electronic Engineering, Imperial College London, London, UK (e-mail: saurab.chhachhi11@imperial.ac.uk; f.teng@imperial.ac.uk)}
\thanks{Digital Object Identifier: }}

\maketitle

\begin{abstract}
    Given the vital role that smart meter data could play in handling uncertainty in energy markets, data markets have been proposed as a means to enable increased data access. However, most extant literature considers energy markets and data markets separately, which ignores the interdependence between them. In addition, existing data market frameworks rely on a trusted entity to clear the market. This paper proposes a joint energy and data market focusing on the day-ahead retailer energy procurement problem with uncertain demand. The retailer can purchase differentially-private smart meter data from consumers to reduce uncertainty. The problem is modelled as an integrated forecasting and optimisation problem providing a means of valuing data directly rather than valuing forecasts or forecast accuracy. Value is determined by the Wasserstein distance, enabling privacy to be preserved during the valuation and procurement process. The value of joint energy and data clearing is highlighted through numerical case studies using both synthetic and real smart meter data.
\end{abstract}

\begin{IEEEkeywords}
data market, smart meters, differential privacy, decision-dependent uncertainty, newsvendor model
\end{IEEEkeywords}

\section{Introduction}
\subsection{Background}
Smart meter data is playing a key role in the energy transition\cite{Wang2019a}. Many novel operating models and pricing schemes, for example, the customised tariff mechanism proposed in \cite{Lai2023}, and the demand response scheme developed in \cite{Ostad2023}, both require access to smart meter data. Most operational uses of smart meter data are connected to an energy retailer's core function of energy procurement. For example, tariff setting\cite{Feng2019} can be viewed as an extension of the retailer energy procurement problem (REPP). Similarly, demand elasticity estimation\cite{Schofield2014} and clustering\cite{Sun2019} can be incorporated as additional pre/post processing steps with the ultimate value derived from their effects on energy procurement decisions. The REPP provides a clear mechanism for how smart meter data quality, through forecast accuracy, affects monetary quantities such as the retailer's profit\cite{Han2020}. Concurrently, it shows that the value of the data is inherently dependent on electricity procurement costs. This interdependence means that the energy and data procurement cannot be considered separately. In addition, digitalisation of the energy sector and the decreasing marginal cost of energy production will shift value increasingly towards handling uncertainty and the data required. Contextual optimisation\cite{Sadana2024} and integrated\cite{Zhou2024} approaches, which incorporate data procurement, uncertainty modelling and decision optimisation, are required to properly assess smart meter data value. 

Most of existing data-driven techniques assume the decision-maker or data user has free access to the required data streams and therefore accrues all the benefits of data access. However, this does not account for the data costs such as privacy concerns\cite{Veliz2018}. In particular, the wealth of personal information embedded within smart meter data mean privacy concerns are of particular importance\cite{Wang2019}. Given that existing data protection regulations in many countries require retailers to actively incentivise consumers to share their data\cite{BEIS2018}, the use of Privacy-Preserving Techniques (PPT) could significantly improve data availability and access. Differential Privacy (DP) is a suitable PPT for smart meter data\cite{Teng2022}. Previous studies have shown that smart meter data with DP protection can provide usable and thus valuable data\cite{Eibl2018}. However, determining the appropriate privacy-utility trade-off requires an integrated approach\cite{Chhachhi2021}. As such, a joint energy and data market framework could facilitate the balancing of access and privacy of smart meter data. 

\subsection{Existing Approaches}
The REPP has been studied extensively with many formulations incorporating a variety of market assumptions (e.g., price-maker/taker) and retailer assumptions (e.g., risk aversion, tariff setting). However, the valuation of data within this context is a relatively new endeavour. An overview of existing generic data market structures can be found in \cite{Chhachhi2024}. Here, we expand on mechanisms proposed specifically for the REPP, in order to compare our proposed valuation and procurement mechanism. 

Ref. \cite{Han2020} proposes a cooperative game framework assuming Gaussian day-ahead load forecasts with scheduled and unscheduled load components. The retailer is able to purchase information on scheduled load to reduce forecast uncertainty. However, the retailer is assumed to be a Trusted Third Party (TTP) with full data access and privacy concerns are not considered. Within this framework the retailer's profit is non-monotonic and therefore does not guarantee individual rationality, even in the absence of privacy concerns. In \cite{Wang2021}, the authors propose the data value rate: a distribution-dependent coefficient of value per unit reduction in the standard deviation of the forecast errors. Consumer reserve prices are assumed to be a function of a privacy sensitivity parameter, which is independent of forecast errors\cite{Wang2022}. The market clearing mechanism assumes perfect competition and therefore does not account for potential incentive compatibility issues. Importantly, both of these methods value forecasts, specifically improvements in forecast uncertainty/accuracy. As such, these indicate the value of a combination of the data and the forecasting model applied rather than the smart meter data itself.

In addition to specific mechanisms proposed for the REPP, there are other closely linked studies. Ref. \cite{Yan2022} proposes a joint energy and data market for a two-stage robust scheduling problem. It assumes the reduction to the uncertainty set offered by the data are known and available to a TTP. As such, privacy and reserve prices are not considered. A similar problem proposed in \cite{Xie2024} does include reserve prices, in the form of prediction costs, however, the privacy of the forecasts in not considered. Ref. \cite{Mieth2023} proposes Distributionally-Robust Optimisation (DRO) based valuation scheme which incorporates privacy concerns through DP. However, it does not consider reserve prices and requires fixed privacy concerns to remain privacy-preserving. Finally, \cite{Zhou2024} proposes an Integrated Forecasting and Optimisation Framework (IFOF) for load data valuation for multi-energy systems. Data value and payments are then calculated using the Shapley value and thus the mechanism suffers from the same drawbacks as \cite{Han2020}. This approach does provide a direct connection between load data and costs. However, it does not jointly optimise the energy and data market costs.

\subsection{Contributions}
Our proposed joint energy and data market framework differs from the existing approaches discussed above in that it focuses on ensuring data privacy, modelling reserve prices, models the effect of differential privacy on retailer profits, values smart meter data directly, and importantly, co-optimises energy and data payments. As such, we make the following contributions:
\begin{itemize}
    \item We develop an IFOF for the REPP allowing for direct data valuation rather than valuations of forecasts/accuracy.
    \item We build upon the joint optimisation mechanism in \cite{Chhachhi2024} to propose a joint energy and data market which co-optimises energy and data procurement costs. The mechanism preserves privacy for both consumer data and the retailer's procurement model, thus eliminating the need for a TTP.
    \item We introduce application-specific calibrations of the mechanism components, including the Lipschitz constant, to improve procurement performance and discuss the theoretical trade-offs involved.
    \item We present multiple numerical case studies using both synthetic data and real smart meter data to validate the performance of the data valuation mechanism against existing approaches that value forecasts, and the data procurement mechanism.
\end{itemize}

The remainder of the paper is organised as follows. Section \ref{sec:theory} presents the proposed joint energy and data market framework derived from the Wasserstein Distance (WD) based task-specific valuation and procurement mechanism presented in \cite{Chhachhi2024}. This is followed by a case study comparing the performance of the proposed mechanism against existing techniques for load forecasts valuation in Section \ref{sec:cs_fore}. Section \ref{sec:cs_sm} details a case study using real smart meter data highlighting the value of jointly optimising energy and data costs. Finally, conclusions are drawn and future research directions discussed in Section \ref{sec:conc}.

\section{Joint Energy and Data Market Framework}\label{sec:theory} 
This sections details the proposed joint energy data market framework. First, we transform the REPP into an IFOF problem, providing a direct relationship between smart meter data and retailer profit. We then outline the mechanism in \cite{Chhachhi2024} and the necessary adaptations required for the REPP, including defining the WDs, the Lipschitz constant, and methods to calibrate risk.

\subsection{Retailer Energy Procurement Problem}\label{sec:ret_prob}
We consider the day-ahead retailer energy procurement detailed in \cite{Han2020}. The retailer procures energy from the (day-ahead) wholesale market, $q$, at a price $\lambda^w$ and receives a fixed rate of $\lambda^r$ for the amount of energy consumed by its consumers. It is assumed that the actual consumer demand, $D$, is a random variable with a PDF, $f_D(y)$, and a CDF, $F_D(y)$. We assume imbalances resulting from forecast errors are settled in a dual price balancing market, with a selling price, $\lambda^{+}$, and buying price, $\lambda^{-}$. We use the same relational assumptions among prices as \cite{Han2020}. 

The aim of the retailer is to maximise their expected profit, $\Pi(q,D)$, by choosing the day-ahead bidding quantity, $q$. This can be expressed as a classical newsvendor problem:
\begin{align}
    \max_q \lambda^{r} E[D] - \lambda^{u}E\left[D-q\right]^{+} + \lambda^{o}E\left[q-D\right]^{+}
\end{align}
where, $\lambda^{o} = \lambda^{w} - \lambda^{+}$ is the overage cost, $\lambda^{o} = \lambda^{-} - \lambda^{w}$ is the underage cost, and $[\cdot]^+$ denotes $\max(\cdot,0)$.

This admits a closed-form solution, $q^* = F^{-1}(\tau)$, where, $\tau = \left(\frac{\lambda^{-} - \lambda^{w}}{\lambda^{-}+\lambda^{+}}\right)$ is the critical fractile. However, this requires knowledge of the true demand distribution, which is not available. Instead, the retailer must estimate the demand distribution, using historical data or through probabilistic forecasting. As a result, the retailer will make a procurement decision, $\hat{q}^*$, based on their estimated demand distribution $\hat{D}$, delivering a sub-optimal profit of $\Pi(\hat{q}^*,D)$. In our context, the retailer would produce day-ahead demand forecasts using reference data, (e.g., national demand or default load profiles) or smart meter data from a subset of its consumers.

\subsection{Integrated Forecasting and Optimisation}
In our framework we aim to value the smart meter data rather than the forecasts produced with it. We therefore develop an explicit modelling framework which allows for the valuation of smart meter data itself.

\subsubsection{Forecasting Model}
Instead of assuming access to the demand forecast distribution, $D$, we consider the scenario where the retailer has access to some historical data $S_{N_s} = \{(d_1,\mathbf{x}_1), \dots, (d_{N_s},\mathbf{x}_{N_s})\}$, where, $d_t$ is the demand and $\mathbf{x}_t$ is a vector of covariates or features (e.g., day of the week, lagged demand values, $d_{t-*}$) \cite{Huber2019}. This data is used to generate probabilistic day-ahead demand forecasts.

The choice of forecasting technique does not affect our framework, as long as the output is Lipschitz w.r.t. the input, $\mathbf{x}$. Here, we focus on the fully-connected feed-forward ANN with a single hidden layer\cite{Huber2019}:
\begin{align}
    \hat{y}(\mathbf{x}) = o(\mathbf{\Psi}^{(2)}a(\mathbf{\Psi}^{(1)}\mathbf{x}+\mathbf{b}^{(1)})+\mathbf{b}^{(2)})
    \label{eq:ann}
\end{align}
where, the inputs nodes $\mathbf{x}$ are connected to the hidden layer nodes via weight matrix, $\mathbf{\Psi}^{(1)}$. The output of the hidden layer is connected to the output layer by $\mathbf{\Psi}^{(2)}$. The bias vectors at each node are $\mathbf{b}^{(1)}$ and $\mathbf{b}^{(2)}$. Finally, $o(\cdot)$ and $a(\cdot)$, represent the activation functions (e.g., ReLu or Sigmoid).

To ensure a particular Lipschitz constant, $K_f$, we limit the norm of the weight matrices, $\lVert \mathbf{\Psi}^{(*)} \rVert \leq K_f$, through a constrained optimisation procedure\cite{Gouk2021}. 

\subsubsection{Optimisation}
We now characterise the resulting forecast distribution and optimal bidding quantity. Disjoint approaches first estimate demand using a forecasting model to produce point forecasts and forecast errors, which are then used to construct a demand distribution\cite{Huber2019}\footnote{For example, SAA, where empirical errors are used directly, or model-based methods, which first estimate distributional parameters.}. These do not provide a direct relationship between the historical input data and the retailer's profit. We adopt an IFOF where the optimal bidding quantity is forecast directly. This approach is equivalent to a quantile regression problem as detailed in Theorem \ref{th:quantile}, below.

\begin{theorem}[Newsvendor Regression]
    The integrated retailer energy forecasting and procurement problem is equivalent to a quantile regression problem for the $\tau$-th quantile, between historical demand data, $d_t$, and a set of features, $\mathbf{x}_t$ (Adapted from \cite[eq. (8)]{Huber2019}):
\begin{align}
    \min_{\mathbf{\Psi}} \sum_{t=1}^{N_s}(\tau-1) \left[d_{t} - q_t(\mathbf{\Psi},\mathbf{x}_t)\right]^+ + \tau \left[q_i(\mathbf{\Psi},\mathbf{x}_t) - d_{t}\right]^+
        \label{eq:jm_quant_reg}
\end{align}
where, $q_t(\mathbf{\Psi},\mathbf{x}_t)$ is the output of the ANN-based forecasting model with weight matrix, $\mathbf{\Psi}$.
\label{th:quantile}
\end{theorem}

\subsection{Joint Optimisation Mechanism}
\begin{figure}[htb]
    \centering
    \includegraphics[width=\linewidth]{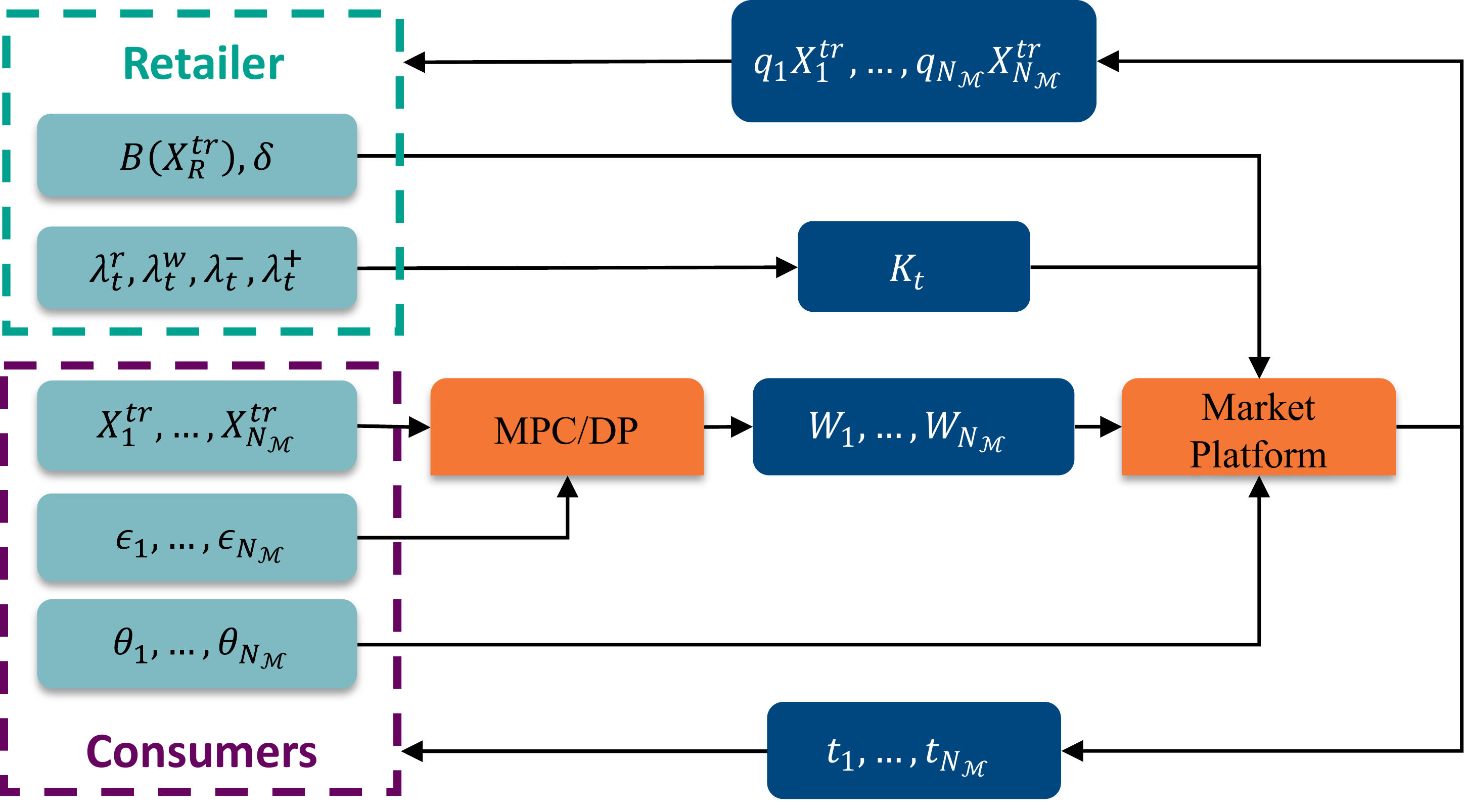}
    \caption{Overview of Joint Optimisation Mechanism}
    \label{fig:framework}
\end{figure}

Having established the retailer problem we now formalise the incentive mechanism, summarised in Figure \ref{fig:framework}. A full detailing of the mechanism can be found in \cite{Chhachhi2024}. Here, we briefly outline the inputs, assumptions, and dataflows.

\subsubsection{Consumers and Data Value}
We assume a scenario where a retailer has a set of consumers, $\mathcal{M}$, each with their own private smart meter data, $X_i$, private reserve price for it, $\theta_i \in \left[\underline\theta_i, \bar\theta_i \right]$, with $0 \leq \underline\theta < \bar\theta < \infty, \quad \forall i \in \mathcal{M}$, and privacy preferences $\epsilon \in \left(0, \infty\right), \quad \forall i \in \mathcal{M}$. We assume each consumer's dataset is a univariate distribution, $X_i^{tr}$, i.e., the training set. The individual value of each consumer's data is defined by the WD: $W_i = W(X_i^{tr},X_T^{tr}) + W(X_i^{DP},\delta^0)$, where, $X_T^{tr} = \frac{1}{N_\mathcal{M}}\sum_{i \in \mathcal{M}} X_i^{tr}$ is the true/target aggregate load, $X_i^{DP}$ is the local-DP noise added to consumer $i$'s data, $\delta^0$ is the Dirac delta distribution, and $N_\mathcal{M} = \lvert \mathcal{M}\rvert$. Only individual WDs need to be calculated as their combinatorial value is approximated by the Hoeffding bound\cite[Theorem 2]{Chhachhi2024}. This is calculated using multi-party computation to preserve consumers' data privacy, and then shared with the market platform. The use of the WD as a valuation metric allows us to capture both the non-I.I.D. nature of the data as well as the effect of DP.

The retailer offers consumers an annual contract consisting of a fixed tariff, $\lambda^r$, and a payment/discount, $t_i$, for data sharing. To determine the discount for each consumer's historical data, $X^{tr}_i$, stored on the smart meter is used. A holdout validation set ($X^{va}_i$) is used to better estimate out-of-sample performance. The true performance is characterised using a test set ($X^{te}_i$)\footnote{We assume a big data regime i.e., $W(X_i^{tr},X_i^{te}) \approx 0, \forall i \in \mathcal{M}$. Discrepancies between the training and test sets can be accounted for within the WD via concentration inequalities\cite{Fournier2015}.}.

\subsubsection{Retailer and Market Platform}
The joint optimisation mechanism in \cite{Chhachhi2024} ensures privacy for both consumers' data and the retailer's model. As such, instead of sharing the REPP, the retailer provides the market platform with the Lipschitz constant of the REPP, $K$, a reference budget/performance, $B(X_R)$, achieved by the retailer in the absence of it's consumers smart meter data, and a confidence level for the Hoeffding bound, $\delta$. 

The IFOF provides a function connecting the input data, $\mathbf{x}$, to the retailer's profit, $\Pi$. In order to obtain the Lipschitz constant for the entire framework we need to consider the Lipschitz constants of each stage resulting in:
\begin{align}
    K = 2K_f \max(\lambda^u,\lambda^o)
\end{align}
where, $K_f$ is the Lipschitz constant of the forecasting model.
Here we note a significant computational advantage of our proposed mechanism over extant cooperative games frameworks. These require training a new model for each time period as the target quantile for newsvendor regression is dependent on market prices. Our approach, on the other hand, only requires calculating the Lipschitz constant.

The retailer is assumed to have access to a publicly available reference dataset, $X_R$, and corresponding reference performance, $B(X_R)$. Ideally, $B(X_R) = \Pi(X_T) - \Pi(X_R)$, however, we do not have access to $\Pi(X_T)$, as this would also violate consumers' data privacy during the procurement process. The retailer is therefore forced to estimate $B(X_R)$, using for example, historical performance, or theoretical problem-specific bounds. We study the effect of under/over-estimation in Section \ref{sec:cs_sm}.

The market platform, which could be the retailer, then aims to joint optimise energy and data procurement, ensuring the smart meter data procured from consumers results in a profit increase greater than the cost of procuring it. It also ensures the retailer's profit is no less than that achieved using solely the freely available reference data, $X_R$. In addition to the information submitted by consumers and the retailer, the platform is assumed to have prior knowledge of the distribution of reserve prices, $\theta$, within the population. The problem is formulated as a mixed-integer second-order cone (MISOCP) with the solution determining the data procurement decisions, $q_i \in \{0,1\}, \forall i \in \mathcal{M}$, and data payments, $t_i \geq 0, \forall i \in \mathcal{M}$. The computational complexity of the MISOCP depends on the assumption placed on the Hoeffding bound approximation. If we consider an infinite population ($INF$) e.g., only a subset of consumers have smart meters or are willing to participate in the market, the resulting MISOCP has $N_\mathcal{M}+1$ binary variables\cite[eq. (27)]{Chhachhi2024}. If, instead, we consider a finite population ($FIN$), i.e., the market includes the full consumer base of the retailer the resulting MISOCP has $N_\mathcal{M}^2 - N_\mathcal{M}$ additional binary variables\cite[eq. (15)]{Chhachhi2024}. The procurement performance of these two variants is explored in detail in Section \ref{sec:cs_sm}.

\subsection{Lipschitz Calibration} \label{sec:lip}
The Lipschitz bound provides a worst-case bound on the performance loss compared to the target distribution. As such, this may result in overly conservative estimates. This is particularly relevant in our setting due to the asymmetric objective\cite{Chhachhi2024}. We propose two methods to calibrate the Lipschitz constant, in such scenarios. The Lipschitz constant is applicable for any distribution. However, problem specific information can significantly tighten this bound (e.g., bounds on demand, restrictions on demand distributions). 

\subsubsection{Transfer Function} A transfer function approach, similar to \cite{Chen2021}, could be adopted.The relationship could be determined using some reference data available to the retailer or be based on simulations. This empirical Lipschitz constant is then:
\begin{align}
    \hat{K} = \max_{i,j} \frac{\Delta\Pi(X_i,X_j)}{W(X_i,X_j)}
\end{align}
where, $X_i,X_j$, are synthetic/reference distributions. 

This method may still lead to conservative estimates of performance loss. Instead, a probabilistic approach using the average ratio between profit difference and the WD could be utilised. However, using these methods prioritises estimation performance over theoretical guarantees, specifically, foregoing guarantees of budget feasibility.

\subsubsection{Lipschitz Relaxation} Alternatively, we can introduce the notion of locally Lipschitz continuity. We limit the input range considered in order to provide a tighter Lipschitz bound. Formally:
\begin{definition}[Locally Lipschitz] A function $f:A\subset \mathbb{R}^n \to \mathbb{R}^m$ is locally Lipschitz at $x_0 \in A$ if there exist constants $\xi>0$ and $K^{\xi}\in \mathbb{R}_+$ such that:
    \begin{align}
        \lvert x - x_0 \rvert \leq \xi \Rightarrow \lvert f(x) - f(x_0) \rvert \leq K^{\xi} \lvert x - x_0 \rvert
    \end{align}
\end{definition}
The improvement attained through this Lipschitz relaxation depends on the demand distribution considered, and the input attribute we choose to limit. 

\paragraph{Example: Gaussian Newsvendor}
To illustrate the effect of the relaxation, we analyse the Gaussian newsvendor problem. If we assume the demand distribution is Gaussian, we obtain a closed-form expression for the retailer's profit\cite{Siegel2021}. The gradient of the profit function is dependent on how far the order quantity is from the mean and the spread of the distribution itself. Proposition \ref{prop:gauss_lip} bounds the profit gradient and resulting locally Lipschitz constant by bounding the difference between bid, $q$, and the optimal bid, $q^*$. Proof has been omitted for brevity.

\begin{proposition}[Locally Lipschitz Constant]
    The Gaussian newsvendor is locally Lipschitz w.r.t. the bidding quantity $q$, if $\lvert q - q^*\rvert \leq \xi$, with a Lipschitz constant:
    \begin{align}
        K_{nv}^{\xi} =  
        \begin{cases}
            \lambda^{u} - \left(\lambda^{u} + \lambda^{o}\right)\Phi\left(\Phi^{-1}(\tau) - \frac{\xi}{\sigma_D}\right), \lambda^u > \lambda^o\\
            \lambda^{o} - \left(\lambda^{u} + \lambda^{o}\right)\Phi\left(-\Phi^{-1}(\tau) - \frac{\xi}{\sigma_D}\right), \lambda^u < \lambda^o
        \end{cases} 
        \label{eq:gauss_lip}
    \end{align}
    \label{prop:gauss_lip}
\end{proposition}

\begin{figure}[htb]
    \centering
    \subfloat[$\frac{\partial\Pi}{\partial q}$\label{fig:gauss_grad}]{
    \includegraphics[width=0.32\linewidth]{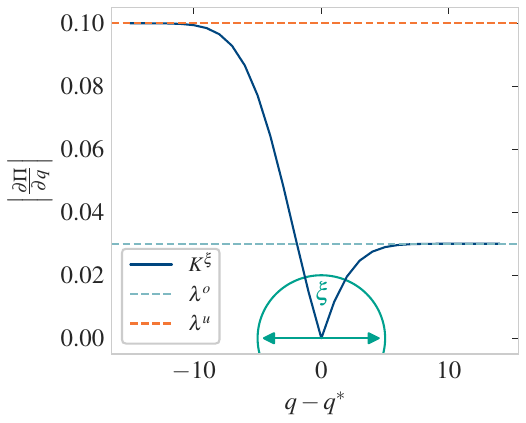}}
    \subfloat[Lipschitz Bound\label{fig:gauss_wd}]{
    \includegraphics[width=0.32\linewidth]{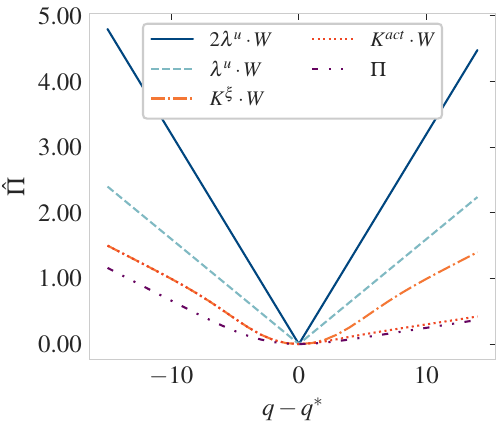}}
    \subfloat[$K^{\xi}_{nv}(\sigma_D)$\label{fig:gauss_sig}]{
    \includegraphics[width=0.32\linewidth]{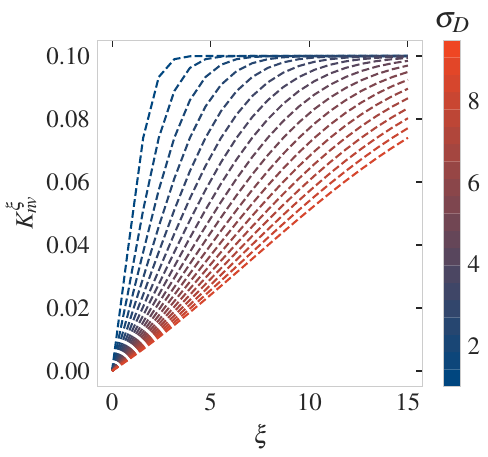}}
    \caption{Lipschitz Relaxation of Gaussian Newsvendor}
    \label{fig:gauss_lip}
\end{figure}

Figure \ref{fig:gauss_grad} shows the change in the gradient as a function of the difference between the bid and the optimal bid, $\lvert q - q^* \rvert$. The performance of the WD-based Lipschitz bounds on the true profit, $\Pi$, using different constants is shown in Figure \ref{fig:gauss_wd}. If we use the global Lipschitz constant, $2\lambda^u$, or even, $\lambda^u$, we obtain a fairly loose, linear bound. If, however, we used the locally Lipschitz constant, $K^{\xi}_{nv}$, we obtain a tighter bound, especially for $q-q^* < 0$, as $\lambda^-$ is higher than $\lambda^+$ in this example. Lastly, if the actual constant, $K^{act}$, were known, a much tighter bound is obtained. The above approach still requires an assessment of the demand uncertainty, $\sigma_D$. Figure \ref{fig:gauss_sig} shows the effect of the underlying demand uncertainty on $K^{\xi}_{nv}$. Increasing the uncertainty reduces the gradient and as a result lowers the constant. As such, we can obtain an upper bound by selecting a theoretically motivated minimum standard deviation (e.g., from national demand or substation forecasts).

Overall, the tightness of the Lipschitz bound is affected, principally by the asymmetry of the cost/profit function of the newsvendor problem. Indeed, if $\lambda^u = \lambda^o$, the problem would be equivalent to median estimation for which the bounds are significantly tighter\cite{Chhachhi2024}.

\section{Case Study: Forecast Procurement} \label{sec:cs_fore}
This case study focuses on comparing the performance of our proposed valuation mechanism, i.e., the use of the WD, against existing mechanisms for load forecast valuation\cite{Wang2021,Han2020,Mieth2023,Jahani2024}. We use the framework presented in \cite{Han2020}, briefly outlined below, and extend it to incorporate the effect of differential privacy.
\subsection{Valuation Framework}
The retailer has $N_\mathcal{M}$ consumers, each with some scheduled load, $L^s_i \sim \mathcal{N}\left(\sum_{i\in \mathcal{M}} \mu^s_i, {\sigma^s}^2_i\right)$, and unscheduled load, $L^u_i \sim \mathcal{N}\left(\sum_{i\in \mathcal{M}} \mu^u_i, {\sigma^u}^2_i\right)$. The retailer produces a reference forecast, $D_{\emptyset}$, consisting of $L^u_i$ and $L^s_i$. The retailer can then purchase actual scheduled load information, $l^s_i$, from it's consumers to reduce forecast uncertainty. The best load forecast, $D_\mathcal{M}$, is achieved when the retailer has scheduled load information from all its consumers. The information procurement is setup as a cooperative game with the value of each coalition, $V(\mathcal{C})$, being determined by the difference in the retailer's expected profit,  $\Pi\left(q^*_\mathcal{C},D_\mathcal{M}\right)$, with, $l^s_i \ \forall i \in \mathcal{C}$, and without scheduled load information, $\Pi\left(q^*_\emptyset,D_\mathcal{M}\right)$. Only coalitions including the retailer generate value. Data payments for each consumer are calculated using Shapley values, $\phi_i$.

To investigate the effect of DP, we assume each consumer has some privacy preference, $\epsilon_i$, and once the retailer purchases the scheduled load information, it will be sent with differentially private noise, using the Gaussian mechanism. The retailer's forecast, $D_\mathcal{C}^{DP}$ with subset, $\mathcal{C}$, of purchased consumers data then has a mean, $\sum_{i\in \mathcal{M}}\mu_i^u+\sum_{i\in \mathcal{M}/\mathcal{C}}\mu_i^s+\sum_{i\in \mathcal{C}} l_i^s$, and variance, $\sum_{i\in \mathcal{M}}{{\sigma_i^u}^2+\sum_{i\in \mathcal{M}/\mathcal{C}}{\sigma_i^s}^2}+\sum_{i\in \mathcal{C}}{\sigma_i^{DP}}^2$. $\sigma_i^{DP}$ is the standard deviation of the DP noise added to $M_i$'s forecast. In addition, we assume $\sum_{i\in \mathcal{M}}{\sigma_i^{DP}}^2=\gamma\sum_{i\in \mathcal{M}}{\sigma_i^s}^2$, where, $\gamma$ is a non-negative noise multiplier.

\subsection{Experimental Setup}
We consider a number of alternative valuation mechanisms to compare the performance of our WD-based mechanism. Their value functions are summarised Table \ref{tab:val_mech}. Within our proposed method, $W_{\mathcal{C},\mathcal{M}}$, we also consider the effect of the finite ($W^{FIN}_{\mathcal{C},\mathcal{M}}$) and infinite ($W^{INF}_{\mathcal{C},\mathcal{M}}$) Hoeffding formulations.

\begin{table}[htb]
    \centering
    \caption{Mechanism Value Functions}
    \label{tab:val_mech}
    \resizebox{\linewidth}{!}{
    \begin{tabular}{r c c}
        \toprule
        Mechanism & V(C) & Source\\
        \midrule
        $\Delta \Pi$ & $\Pi\left(q_\mathcal{C}^*,D_\mathcal{M}\right) - \Pi\left(q_\emptyset,D_\mathcal{M}\right)$ & \cite{Han2020}\\
        $\Delta \sigma$ & $C_\sigma \left(\sigma_{D_\emptyset} - \sigma_{D_\mathcal{C}}\right)$ & \cite{Wang2021}\\
        $DRO_{\mathcal{C},\mathcal{M}}$\textsuperscript{a} & $\Pi\left(q_\mathcal{C}^*,D_\mathcal{C}\right) -\lambda^uW\left(D_\mathcal{C},D_\mathcal{M}\right) - \Pi\left(q_\emptyset^*,D_\mathcal{M}\right)$ & \cite{Mieth2023}\\
        $W_{\mathcal{C},\emptyset}$ & $K\cdot W\left(D_{\emptyset},D_\mathcal{C}\right)$ & \cite{Jahani2024}\\
        $W_{\mathcal{C},\mathcal{M}}$\textsuperscript{b} & $\left[ \Pi(q_\mathcal{M}^*,D_\mathcal{M}) - \Pi(q_\emptyset^*,D_\mathcal{M}) - K\cdot W\left(D_\mathcal{M},D_\mathcal{C}\right) \right]^+$ & \\
        \bottomrule
        \multicolumn{3}{p{\linewidth}}{Notes: \textsuperscript{a} Uses solution to DRO REPP\cite{Lee2021}. \textsuperscript{b}We use the zero-Shapley policy\cite{Liu2020} as negative coalition values are not necessarily an indication of actual reduction in profit but may be a reflection of the conservatism of the approach.}
    \end{tabular}
    }
\end{table}

For the extended model, which incorporates DP, we consider two methods to model the effect of DP\cite{Chhachhi2023}:
\begin{itemize}
    \item $W^{DP}$: Exact expression for the forecast variance, which assumes the data is known to be Gaussian.
    \item $W^{UB}$: Upper bound on forecast variance, with no distributional assumption.
\end{itemize}
In the original model consumers data value is related to the proportion of schedulable load they have\cite{Han2020}. In addition, as shown in \cite{Chhachhi2024}, correlations between privacy parameters and value impact performance. We consider four privacy scenarios of how the total DP noise, $\sum_{i\in \mathcal{M}}{\sigma_i^{DP}}^2$, is distributed among consumers:
\begin{enumerate}
    \item Corr: Proportional to schedulable load.
    \item Inv: Inversely proportional to schedulable load.
    \item Uni: Allocated uniformly.
    \item Rand: Allocated randomly.
\end{enumerate}

We use the same market prices and data generation method as in \cite{Han2020}. However, we use the modified smart meter dataset from \cite{Chhachhi2021} to produce the 8 half-hourly consumer load profiles. For the DP scenarios we first assume $\gamma = 0.5$. As discussed in Section \ref{sec:lip}, the looseness of the Lipschitz bound in our setting can result in conservative value estimates. Without appropriate calibration, the WD-based mechanism with the global Lipschitz constant, as defined in (\ref{eq:jm_lag}), produces a high proportion (up to 80\%) of non-positive coalitions. Therefore, the main results are presented using a calibrated Lipschitz constant. Specifically, we use the reference forecast resulting in $K=\frac{\Pi\left(q_\mathcal{M}^*,D_\mathcal{M}\right) -\Pi\left(q_\emptyset^*, D_\mathcal{M}\right)}{W(D_\emptyset,D_\mathcal{M})}$. The implications of calibration on the risk of budget infeasibility are studied in more detail in Section \ref{sec:cs_sm}.

\subsection{Results}
\subsubsection{Forecast Valuation}
To evaluate the performance of the different valuation methods we compare them without considering DP. The performance metrics, which include correlations to the actual profit difference ($\Delta\Pi$) and the associated the Shapley mis-allocations, are summarised in Table \ref{tab:fc_perf}. First, we see that the WD-based approaches and $DRO_{\mathcal{C},\mathcal{M}}$ are highly correlated with the actual profit, with our proposed method giving a correlation coefficient of $\rho(\Delta\Pi,W_{\mathcal{C},\mathcal{M}})=0.99$\footnote{Without calibration the corresponding correlation is $\rho(\Delta\Pi,W_{\mathcal{C},\mathcal{M}})=0.88$.}. The data value rate ($\Delta\sigma$) performs significantly worse as it only considers changes in the standard deviation of the forecast but not the mean\footnote{If $l_i^s = \mu_i^s \forall i \in \mathcal{M}$, $\Delta\sigma$ and $W_{\mathcal{C},\emptyset}$ would be equivalent up to a constant factor.}. $\Delta\sigma$ ensures non-negativity of coalition values, however, it forgoes budget feasibility ($\Delta \Phi \geq 0$). Conversely, $W_{c,\mathcal{M}}$ and $W^{FIN}_{\mathcal{C},\mathcal{M}}$ ensure budget balance while  $W^{INF}_{\mathcal{C},\mathcal{M}}$ leads to a budget feasible but not balanced allocation. 

Our proposed valuation mechanism, $W_{\mathcal{C},\mathcal{M}}$, results in the lowest mis-allocation ($\Delta \lvert\Phi\rvert$), both in terms of £ and \%, compared to using, $\Delta\Pi$. Although $DRO_{\mathcal{C},\mathcal{M}}$ also performs well, it requires the calculation of the profit and WD of each coalition. In addition, $W_{\mathcal{C}, \emptyset}$ does not provide theoretical guarantees of obtaining a better forecast, only that it will be different from the reference forecast $D_\emptyset$. The Hoeffding approximations provide computational improvements but do reduce performance in terms of mis-allocations.
	
\begin{table}[htb]
    \centering
    \caption{Mechanism Performance Metrics}
    \resizebox{\linewidth}{!}{
    \begin{tabular}{rcccccc}
    \toprule
    & $\Delta\sigma_{\mathcal{C},\emptyset}$ & $DRO_{\mathcal{C},\mathcal{M}}$ &  $W_{\mathcal{C},\emptyset}$ &  $W_{\mathcal{C},\mathcal{M}}$ &  $W^{FIN}_{\mathcal{C},\mathcal{M}}$ &  $W^{INF}_{\mathcal{C},\mathcal{M}}$ \\
    \midrule
    $\rho(\Delta\Pi,*)$ & 0.46 & 0.94 & 0.96 & \textbf{0.99} & 0.83 & 0.86 \\
    $\leq0$\textsuperscript{1} & \textbf{0.00} & 0.58 & \textbf{0.00} & 0.20 & 0.04 & 0.05 \\
    $\Delta \Phi$\textsuperscript{2} & 0.39 & \textbf{0.00} & \textbf{0.00} & \textbf{0.00} & \textbf{0.00} & -0.53 \\
    $\Delta \lvert\Phi\rvert (\pounds)$\textsuperscript{3}& 1.42 & 0.36 & 0.25 & \textbf{0.24} & 0.83 & 0.84 \\
    $\Delta \lvert\Phi\rvert (\%)$\textsuperscript{3} & 0.02 & 0.03 & \textbf{0.01} & \textbf{0.01} & 0.02 & 0.02 \\
    \bottomrule
    \multicolumn{7}{p{\linewidth}}{Notes: \textsuperscript{1}Proportion of coalition values which are non-positive. \textsuperscript{2}Total value difference, $\Delta\Phi = \sum_i \phi^*_i - \sum_i \phi^{\Delta\Pi}_i$. \textsuperscript{3}Shapley mis-allocations, $\Delta \lvert\Phi\rvert = \sum_i \lvert \phi^{\Delta\Pi}_i - \phi^*_i \rvert$.}
\end{tabular}}
    \label{tab:fc_perf}
\end{table}
\subsubsection{Effect of Differential Privacy}
We now consider the addition of DP. The noise introduced into the forecast leads to increased costs compared to the non-DP scenario. As a result, the total payoffs in £ will be less. Given the imposed relationship between $\sum_i\sigma_i^{DP}$ and $\sum_i\sigma_i^s$ the overall value reduction is the same in each privacy setting, only the distribution across consumers varies. For example, when $\gamma=0.5$ the profit reductions observed are 32\% and 60\% using the exact and upper bound formulations, respectively. 
\begin{figure*}
    \centering
    \subfloat[$M_1$\label{fig:del_0}]{\includegraphics[width=0.25\textwidth]{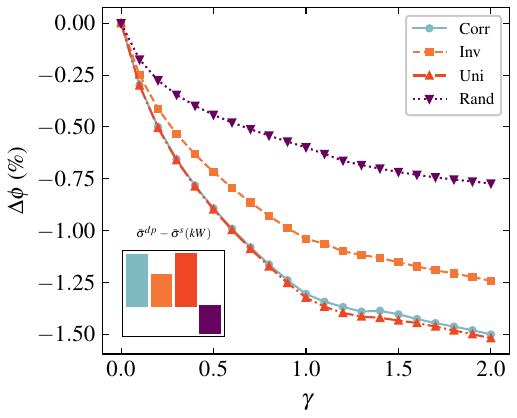}} \hfil
    \subfloat[$M_2$\label{fig:del_1}]{\includegraphics[width=0.25\textwidth]{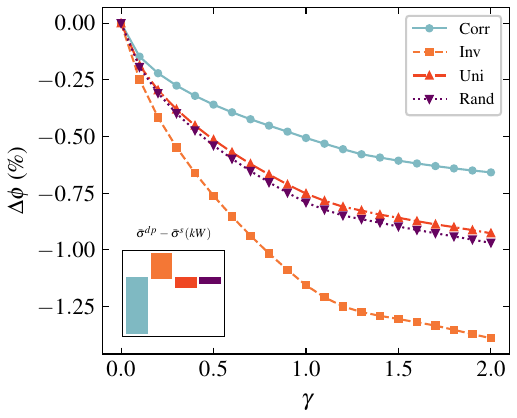}} \hfil
    \subfloat[$M_3$\label{fig:del_2}]{\includegraphics[width=0.25\textwidth]{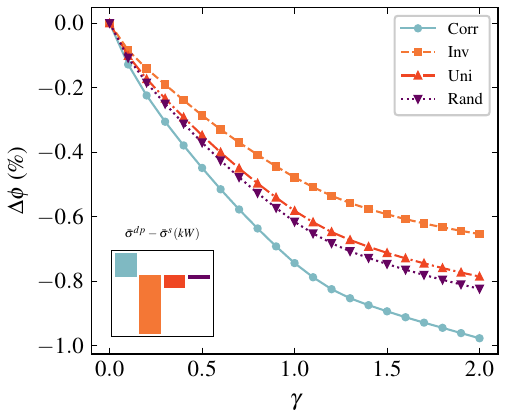}}\hfil
    \subfloat[$M_4$\label{fig:del_3}]{\includegraphics[width=0.25\textwidth]{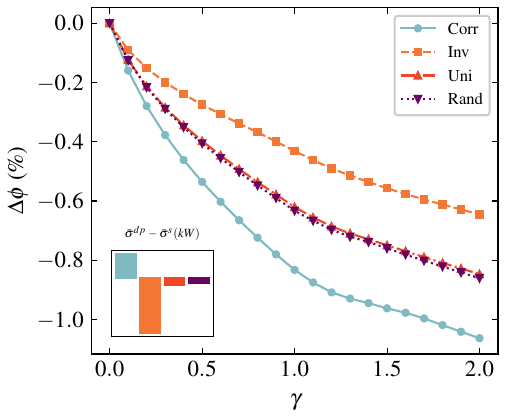}}\\
    \subfloat[Correlated (Corr)\label{fig:sweep_corr}]{\includegraphics[width=0.25\textwidth]{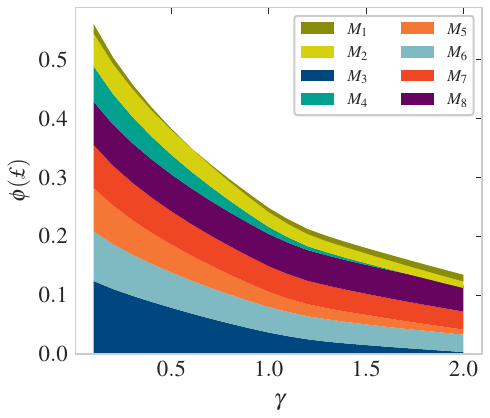}} \hfil
    \subfloat[Inverse (Inv)\label{fig:sweep_inv}]{\includegraphics[width=0.25\textwidth]{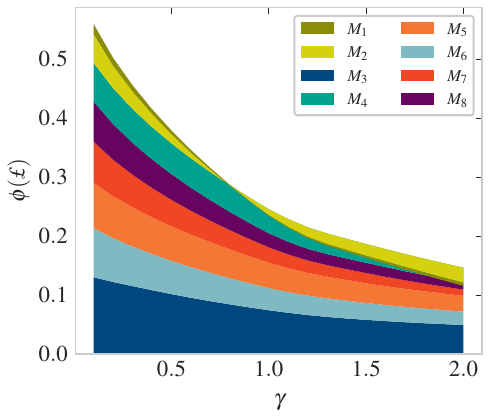}} \hfil
    \subfloat[Uniform (Uni)\label{fig:sweep_uni}]{\includegraphics[width=0.25\textwidth]{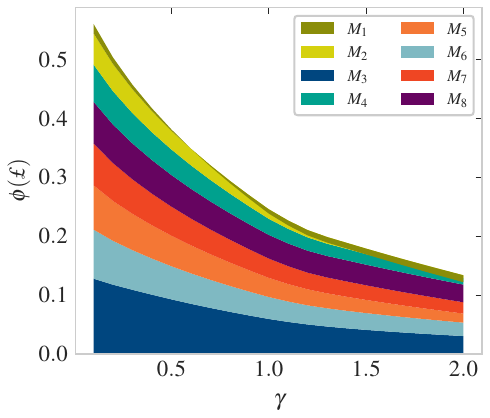}}\hfil
    \subfloat[Random (Rand)\label{fig:sweep_rand}]{\includegraphics[width=0.25\textwidth]{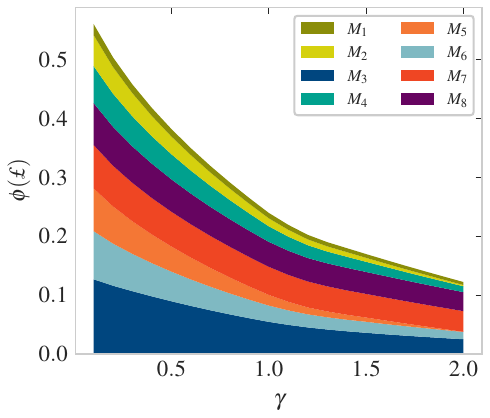}}
    \caption{Effect of Differential Privacy on Forecast Value. Top: Percentage Change in Shapley Allocation and $\sigma^{dp}-\sigma^{s}$, under $W^{DP}$, for Four consumers. Bottom: Shapley Payments with Increasing Noise, $\gamma$, under Different Privacy Scenarios.}
    \label{fig:dp_scens} 
\end{figure*}

Comparing the Shapley allocations (not shown), we find that with $W^{DP}$ the main driver of value remains the average schedulable load, as they are in line with the non-DP scenario. The relative rankings of consumers are preserved across the DP correlation scenarios. The impact of DP noise is more prominent when using $W^{UB}$. This results in zero or negative allocations for some consumers, as their contribution to improving mean forecast is less than the introduced noise.

Changes in allocations across the different correlation scenarios depend on the average schedulable load, the difference in the scheduling uncertainty, $\sigma^s$ and the DP noise, $\sigma^{dp}$, as well as the proportion of positive coalition values. Figure \cref{fig:dp_scens}\cref{fig:del_0}-\cref{fig:del_3} show the percentage change in the Shapely allocations for each consumer as a function of $\gamma$, when using $W^{DP}$. The inset in the bottom left of each plot shows $\sigma^s-\sigma^{dp}$ for each consumer, elucidating the role it plays in determining the changes in allocations. For example, we see that for $M_1$, its' $\sigma^s-\sigma^{dp}$ is lowest in the random correlation scenario, followed by the inverse scenario, with the uniform and correlated scenario resulting in similar values. We see this ordering reflects directly in the change in payoffs as $\gamma$ increases.

We now consider the absolute changes as a function of $\gamma$, as shown in Figures \cref{fig:dp_scens}\cref{fig:sweep_corr}-\cref{fig:sweep_rand}. Value reductions, with an increase in $\gamma$, follow the expected DP correlation scenarios. Specifically, value reduction is most prominent for highest value consumers ($M_3$) in correlated case, with the opposite observed in the inverse case consumers ($M_1$ \& $M_2$). In the uniform and random case, value reduction is more evenly spread. We also see that in the inverse case, in particular, as we increase $\gamma$ past 1 the value of lowest valued consumers becomes negative.

Alongside the overall reduction in value, we also observe that as privacy concerns increase (i.e., $\gamma$ increases), the retailer captures more of the value that remains. This is due to the retailers pivotal role in producing positive coalition values, which becomes more pronounced as overall value decreases. Effectively, the contribution from consumers data is diminished due to the increased noise, while the retailers contribution remains fixed. We see that this effect is more prominent for $W^{DP}$ with the retailer's proportion of value capture increasing from 46\% for $\gamma=0$ to 67\% for $\gamma=2$, whereas for $W^{UB}$ it only increases to 56\%.

\section{Case Study: Procuring Smart Meter Data} \label{sec:cs_sm}
The previous case study valued demand forecasts, rather than smart meter data. As a result, data value was a function of both the underlying smart meter data and the forecasting model used. In this case study we apply the IFOF, for the REPP, to value real smart meter data directly. This more accurately models the scenario faced by retailers who need to actively incentivise consumers to share smart meter data\cite{Teng2022}.

\subsection{Experimental Setup}
We use the modified smart meter dataset from \cite{Chhachhi2021}, which includes half-hourly data for 6010 smart meters spanning 75 weeks, clustered into 8 consumer archetypes. Data is split into a training, validation and test set with a 70-10-20 split. We assume $X_R$ is the GB national demand re-scaled to coincide with $X_T$. The ANN forecasting models are implemented in Python using skforecast and Keras\footnote{Hyper-parameters for all models used are based on a backtesting-based grid search of a mean forecaster with the $X_T^{tr}$(20 epochs, $10^-3$ learning rate, 3 neurons hidden layer, ADAM optimiser).}. We include weight clipping constraints to ensure $K_f = 1$. The input data for the ANN is:
\begin{align}
    \mathbf{x}_t = [X_{-h}, X_{-h-1}, X_{-2h}, X_{-2h-1}, X_{-3h}, X_{-3h-1}]
    \label{eq:jm_lag}
\end{align}
where, $X_{-*}$ are the lagged/historical load values, h is the number of periods in the day (48).

To evaluate the performance of the proposed framework ($K\cdot W$), we first compare its data valuations to the true data value, represented by the profit differences ($\Delta\Pi$), and extant mechanisms which rely on arbitrary metrics ($\Delta RMSE$ and $\Delta MAE$) as proxies for data value \cite{Goncalves2020}. We then examine the impact of model mis-specification by altering the lag structure in (\ref{eq:jm_lag})\footnote{$\mathbf{x}_t = [X_{-h+6}, X_{-h+10}, X_{-2h+6},$ $ X_{-2h+10}, X_{-3h-10}, X_{-3h-4}]$}. We then proceed to analyse the procurement performance of produced by the proposed framework under the finite, $FIN$, and infinite, $INF$, population assumption. We simulate consumers' reserve prices over 50 trials with $\theta \sim U(0,\bar\theta)$, where $\bar\theta \in \left[ 0, 0.1\frac{B(X_R)}{N_\mathcal{M}},\dots, 4\frac{B(X_R)}{N_\mathcal{M}}\right]$\footnote{Due to the large number of simulations that would be required we omit the effect of DP from this case study.}. Finally, we consider the trade-offs between guarantees of budget feasibility and the retailer's profit through sensitivity analyses of the Hoeffding bound confidence level, $\delta \in \left(0,1\right)$, the Lipschitz constant, $K \in \left(0, 2\max(\lambda^u, \lambda^o\right)$, and the reference budget, $B(X_R) \in \left[0, K\cdot W(X_R,X_T)\right]$.

\subsection{Results}
\begin{figure*}
    \centering
    \subfloat[$\Pi(X_c)$ by Valuation Metric\label{fig:jm_proc_metric}]{\includegraphics[width=0.28\textwidth]{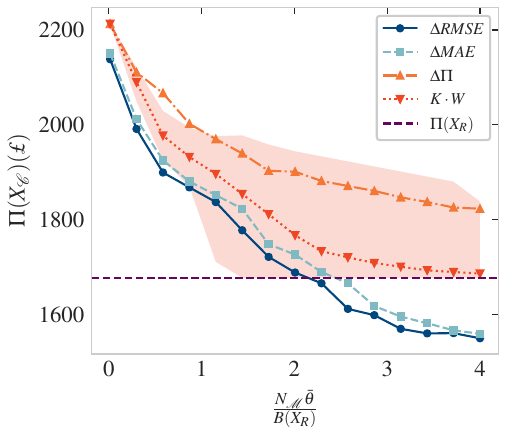}} \hfil
    \subfloat[$\Pi(X_c)$ by Procurement Mechanism\label{fig:jm_proc_cen}]{\includegraphics[width=0.28\textwidth]{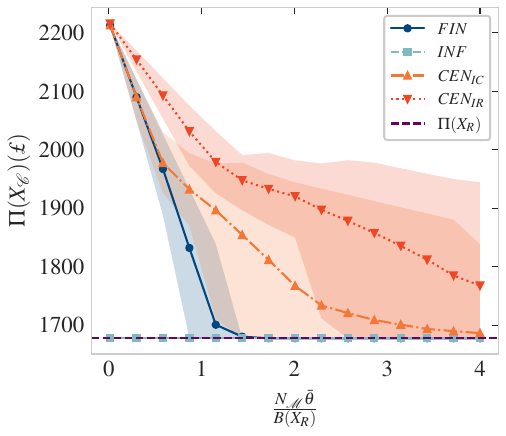}} \hfil
    \subfloat[Data Payments under $FIN$\label{fig:jm_cost_fin}]{\includegraphics[width=0.3\textwidth]{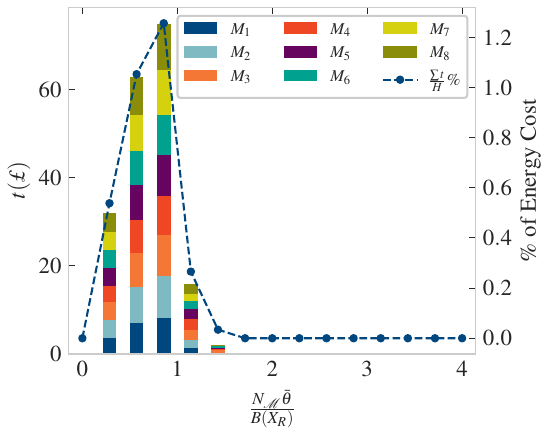}}\\
    \subfloat[$\delta \left(\sfrac{N_\mathcal{M} \bar\theta}{B(X_R)} = 1.44\right)$ \label{fig:jm_risk_hoef}]{\includegraphics[width=0.33\textwidth]{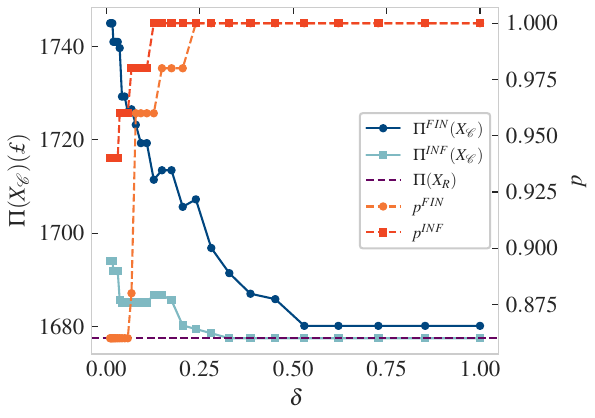}} \hfil
    \subfloat[$K \left(\sfrac{N_\mathcal{M} \bar\theta}{B(X_R)} = 1.15\right)$\label{fig:jm_risk_lip}]{\includegraphics[width=0.33\textwidth]{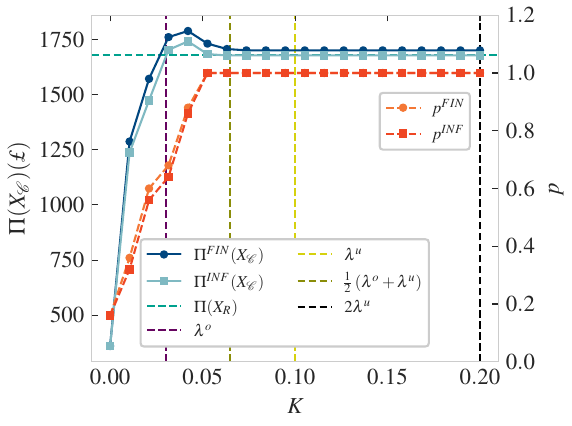}} \hfil
    \subfloat[$B(X_R) \left(\sfrac{N_\mathcal{M} \bar\theta}{B(X_R)} = 1.44\right)$\label{fig:jm_risk_ref}]{\includegraphics[width=0.33\textwidth]{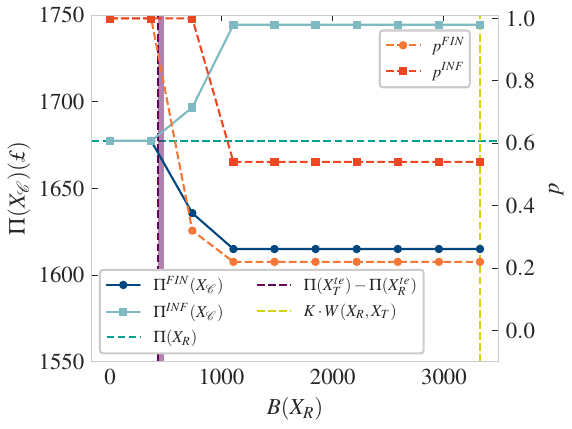}}
    \caption{Valuation of Real Smart Meter Data $(\rho=0)$. Top: Base Results for Proposed Mechanism $(\delta = 0.95, K = 2\lambda^u, B(X_R) = \Pi(X_T^{va}) - \Pi(X_R^{va}))$. Bottom: Effect of Risk Adjustment/Calibration of Conservatism by varying $\delta$, $K$ and $B(X_R)$.}
\end{figure*}

\subsubsection{Data Valuation Metric}
We start by analysing the correlations between the true value function, $\Delta\Pi$, and the other value functions considered. We observe that the $\Delta RMSE$, $\Delta MAE$, and $K \cdot W$ exhibit similar correlation coefficients of between 0.82 and 0.87 across the dataset. However, by calibrating $K$ we can significantly increase the correlations to between 0.88 and 0.92 suggesting that the WD is better able to capture the value dynamics than $\Delta RMSE$ or $\Delta MAE$. Similarly, the WD-based valuations provide more accurate Shapley allocations.
Next, we consider the model mis-specification and the use of reference data instead of the actual smart meter data. Both cases can lead to significant profit reductions. In this case, a $\sim$25\% and a $\sim$5\% reduction in profit are observed when using the national demand and the mis-specified model, respectively, with the later also affecting Shapley allocations. This highlights an advantage of our mechanism which is immune to model manipulation as it relies only on differences in input data.

\subsubsection{Retailer Profit}
We now evaluate the procurement mechanism. Figure \ref{fig:jm_proc_metric}, shows the retailer profit assuming access to all coalition values, across different valuation metrics as reserve prices, $\bar\theta$, increase. $\Delta \Pi$ provides a benchmark of the best-case performance if the retailer knew the profit achieved by each coalition. Conversely, $\Pi(X_R)$ provides the worst-case profit the retailer achieves if they used the reference data. Our approach, $K \cdot W$, performs similar to $\Delta \Pi$ for lower reserve prices and remains better than the other potential metrics for larger reserve prices, selecting higher valued coalitions. Importantly, our approach, unlike $\Delta RMSE$ or $\Delta MAE$, maintains budget feasibility, as shown by the shaded region representing the profit range across all 50 trials. 

Figure \ref{fig:jm_proc_cen} shows the performance of our proposed procurement mechanisms, $FIN$ and $INF$. We see that the finite formulation procures data for $\frac{N_\mathcal{M}\ \bar{\theta}}{B(X_R)}\leq 1.5$, whereas the infinite formulation is too conservative. One reason for this conservatism is the Hoeffding bound. Indeed, the centralised equivalents ($CEN_{IC}$,$CEN_{IR}$)\footnote{$CEN_{IR}$ ensures individual rationality while $CEN_{IC}$ ensures incentive compatibility. See \cite[Section 4.2]{Chhachhi2024} for details.} perform much better, offering increased profits across the reserve price range considered. 

\subsubsection{Consumer Data Payments}
Next, we turn to the data payments made to consumers. Figure \ref{fig:jm_cost_fin} shows the average annualised data payments made to the 8 consumers for $FIN$. We see that as consumers' reserve prices increase, their payments initially increase before reducing. This is because in the initial phase the budget constraints, on average, do not affect the procurement decisions. As a result, the retailer can continue to procure data even as the compensation demanded by consumers increases. However, at a certain point the budget constraints start to limit the feasible coalitions that can be purchased. The annual average total data payments range between £2 and £75 which represent up to 1.25\% of the total energy costs ($H$). However individual proportions can be much higher with, for example, $M_5$ receiving a 16.24\% discount for sharing data under particular runs.

\subsubsection{Risk and Calibration}
We now investigate the effect of risk adjustment and calibration on the retailer's profit. First, we vary the confidence level, $\delta$, of the Hoeffding bound. As shown in Figure \ref{fig:jm_risk_hoef}, reducing $\delta$, results in increased profits when $\delta <0.5$ for both $FIN$ and $INF$. However, this comes with an increased probability of budget infeasibility, $p$, which is plotted on the right-hand axis. For example, we see that by reducing $\delta$ to 0.06, the average profit increases by 2.73\% but $p$ reduces to 0.86. However, that there is a significant range ($\delta$ = 1 to 0.24) for which budget feasibility is not affected, suggesting our underlying valuation is conservative.

In Figure \ref{fig:jm_risk_lip} we consider the relaxation of the Lipschitz constant, $K$. If $K$ is too small, the valuation mechanism over-estimates the potential improvement, and over-purchases data. As a result, for both $FIN$ and $INF$, the profits, $\Pi(X_\mathcal{C})$, are lower than the reference profit, $\Pi(X_R)$. However, if $K$ is between $\lambda^o$ and $\frac{1}{2}\left(\lambda^o + \lambda^u\right)$, we see an improvement in the resulting profit. This, again, comes with a risk of budget infeasibility, with $p$ between 0.6 and 1. As $K$ is increased beyond $\frac{1}{2}\left(\lambda^o + \lambda^u\right)$, the mechanisms revert to the subset procured when using the global Lipschitz constant, $2\lambda^u$, and budget feasibility is maintained.

Finally, we consider the effect of the reference budget estimate, as shown in Figure \ref{fig:jm_risk_ref}. While $B(X_R) \leq \Pi(X_T^{te}) - \Pi(X_R^{te})$, budget feasibility is maintained. Increasing $B(X_R)$ beyond this can lead to either an increase in profits, as is the case with $INF$, or a decrease in overall profits in the case of $FIN$. When the base case procurement decision was overly conservative ($INF$), increasing the budget reference allows more data to be purchased while potentially maintaining budget feasibility. We see a 4\% improvement in profit but this comes at the cost of $p$ reducing to 0.54. Conversely, when using a less conservative valuation approach ($FIN$), increasing the budget leads to significant over-procurement, on average leading to reduced profits and reduced $p$. 

Overall, we find that over-estimation of $B(X_R)$ has the largest impact on budget feasibility, suggesting the Hoeffding confidence level, $\delta$ and the Lipschitz constant, $K$, are more suitable tools to calibrate risk levels.

\section{Conclusion}\label{sec:conc}
In this paper we develop the concept of a joint energy and data market for differentially-private smart meter data. This is achieved by proposing a data valuation mechanism based on the WD and a data procurement mechanism leveraging incentive mechanism design theory to derive a tractable MISOCP formulation. Two case studies were presented to demonstrate the efficacy of the proposed mechanism. The first case study considered the procurement of forecasts while the second considered the procurement of smart meter data directly. The first highlighted the superior performance of the WD over existing methods for data valuation. The second implemented the integrated approach which ensures budget feasibility by capturing the inherent decision-dependent structure in such a market, in contrast to arbitrary data utility metrics, such as the RMSE, proposed in extant literature. Finally, risk calibration of the proposed framework was discussed and validated experimentally through sensitivity analyses.

The joint energy and data market framework proposed in this paper focuses on ensuring privacy preservation, for both consumer data and the retailer's model, and theoretical guarantees of budget feasibility. As such, it introduces levels of approximation (Hoeffding and Lipschitz bounds) which contribute to the conservatism of the approach. Future work will focus on introducing methods to improving the balance between valuation accuracy, computational tractability and theoretical guarantees. This includes the development of probabilistic Lipschitz bounds and Wasserstein geodesics for improved combinatorial accuracy. In addition, the framework will be extended to a multi-buyer environment to more accurately reflect options available to consumers, and to model demand response/tariff setting problems. 

\bibliographystyle{IEEEtran}
\bibliography{references}

\begin{thebibliography}{10}
\providecommand{\url}[1]{#1}
\csname url@samestyle\endcsname
\providecommand{\newblock}{\relax}
\providecommand{\bibinfo}[2]{#2}
\providecommand{\BIBentrySTDinterwordspacing}{\spaceskip=0pt\relax}
\providecommand{\BIBentryALTinterwordstretchfactor}{4}
\providecommand{\BIBentryALTinterwordspacing}{\spaceskip=\fontdimen2\font plus
\BIBentryALTinterwordstretchfactor\fontdimen3\font minus \fontdimen4\font\relax}
\providecommand{\BIBforeignlanguage}[2]{{%
\expandafter\ifx\csname l@#1\endcsname\relax
\typeout{** WARNING: IEEEtran.bst: No hyphenation pattern has been}%
\typeout{** loaded for the language `#1'. Using the pattern for}%
\typeout{** the default language instead.}%
\else
\language=\csname l@#1\endcsname
\fi
#2}}
\providecommand{\BIBdecl}{\relax}
\BIBdecl

\bibitem{Wang2019a}
Y.~Wang, Q.~Chen \emph{et~al.}, ``Review of smart meter data analytics: Applications, methodologies, and challenges,'' \emph{IEEE Trans. Smart Grid}, vol.~10, pp. 3125--3148, 5 2019.

\bibitem{Lai2023}
S.~Lai, J.~Qiu \emph{et~al.}, ``Customized pricing strategy for households based on occupancy-aided load disaggregation,'' \emph{IEEE Trans. Energy Markets, Policy and Regulat.}, vol.~1, pp. 118--130, 6 2023.

\bibitem{Ostad2023}
M.~Ostadijafari, G.~Manandhar \emph{et~al.}, ``Principal-agent model for bilateral contract design to incentivize residential demand-side flexibility,'' \emph{IEEE Trans. Energy Markets, Policy and Regulat.}, vol.~1, pp. 310--321, 12 2023.

\bibitem{Feng2019}
C.~Feng, Y.~Wang \emph{et~al.}, ``{Smart Meter Data-Driven Customizing Price Design for Retailers},'' \emph{IEEE Trans. Smart Grid}, 2020.

\bibitem{Schofield2014}
J.~Schofield, R.~Carmichael \emph{et~al.}, ``Report a3 low carbon london learn. lab - residential consumer responsiveness to time-varying pricing,'' 2014.

\bibitem{Sun2019}
M.~Sun, Y.~Wang \emph{et~al.}, ``Clustering-based residential baseline estimation: A probabilistic perspective,'' \emph{IEEE Trans. Smart Grid}, vol.~10, pp. 6014--6028, 11 2019.

\bibitem{Han2020}
L.~Han, J.~Kazempour \emph{et~al.}, ``Monetizing customer load data for an energy retailer: A cooperative game approach,'' in \emph{2021 IEEE PowerTech}, 2021, pp. 1--6.

\bibitem{Sadana2024}
U.~Sadana, A.~Chenreddy \emph{et~al.}, ``A survey of contextual optimization methods for decision-making under uncertainty,'' \emph{Eur. J. Oper. Res.}, 2024.

\bibitem{Zhou2024}
Y.~Zhou, Q.~Wen \emph{et~al.}, ``Load data valuation in multi-energy systems: An end-to-end approach,'' \emph{IEEE Trans. Smart Grid}, pp. 1--1, 2024.

\bibitem{Veliz2018}
C.~Véliz and P.~Grunewald, ``Protecting data privacy is key to a smart energy future,'' \emph{Nature Energy}, vol.~3, pp. 702--704, 9 2018.

\bibitem{Wang2019}
Y.~Wang, Q.~Chen \emph{et~al.}, ``Deep learn.-based socio-demographic inf. identification from smart meter data,'' \emph{IEEE Trans. Smart Grid}, vol.~10, pp. 2593--2602, 5 2019.

\bibitem{BEIS2018}
BEIS, ``Smart metering implementation programme: Review of the data access and privacy framework,'' BEIS, Tech. Rep., 2018.

\bibitem{Teng2022}
F.~Teng, S.~Chhachhi \emph{et~al.}, ``Balancing privacy and access to smart meter data: an {E}nergy {F}utures {L}ab briefing paper,'' \emph{Imperial College London}, pp. 1--64, 5 2022.

\bibitem{Eibl2018}
G.~Eibl, K.~Bao \emph{et~al.}, ``The influence of differential privacy on short term electric load forecasting,'' \emph{Energy Informatics}, 2018.

\bibitem{Chhachhi2021}
S.~Chhachhi and F.~Teng, ``Market value of differentially-private smart meter data,'' in \emph{2021 IEEE PES Innovative Smart Grid Techno. Conf.}\hskip 1em plus 0.5em minus 0.4em\relax IEEE, 2 2021, pp. 1--5.

\bibitem{Chhachhi2024}
------, ``Wasserstein markets for differentially-private data,'' \emph{ar{X}iv}, 12 2024.

\bibitem{Wang2021}
Y.~Wang, I.~L. Bennani \emph{et~al.}, ``Electricity consumer characteristics identification: A federated learning approach,'' \emph{IEEE Trans. Smart Grid}, vol.~12, pp. 3637--3647, 7 2021.

\bibitem{Wang2022}
B.~Wang, Q.~Guo \emph{et~al.}, ``Mechanism design for data sharing: An electricity retail perspective,'' \emph{Appl. Energy}, vol. 314, p. 118871, 5 2022.

\bibitem{Yan2022}
M.~Yan and F.~Teng, ``Towards joint electricity and data trading: A scalable cooperative game theoretic approach,'' \emph{ar{X}iv}, 10 2022.

\bibitem{Xie2024}
R.~Xie, P.~Pinson \emph{et~al.}, ``Robust generation dispatch with purchase of renewable power and load predictions,'' \emph{IEEE Trans. Sustain. Energy}, pp. 1--16, 2024.

\bibitem{Mieth2023}
R.~Mieth, J.~M. Morales \emph{et~al.}, ``Data valuation from data-driven optimization,'' \emph{IEEE Trans. Control Netw. Sys.}, pp. 1--12, 2024.

\bibitem{Huber2019}
J.~Huber, S.~Müller \emph{et~al.}, ``A data-driven newsvendor problem: From data to decision,'' \emph{Eur. J. Oper. Res.}, vol. 278, pp. 904--915, 11 2019.

\bibitem{Gouk2021}
H.~Gouk, E.~Frank \emph{et~al.}, ``Regularisation of neural networks by enforcing lipschitz continuity,'' \emph{Machine Learn.}, vol. 110, pp. 393--416, 2 2021.

\bibitem{Fournier2015}
N.~Fournier and A.~Guillin, ``On the rate of convergence in wasserstein distance of the empirical measure,'' \emph{Probability Theory and Related Fields}, vol. 162, pp. 707--738, 8 2015.

\bibitem{Chen2021}
L.~Chen, Z.~Wu \emph{et~al.}, ``oward future information market: An information valuation paradigm,'' in \emph{2021 IEEE PES General Meeting}.\hskip 1em plus 0.5em minus 0.4em\relax IEEE, 7 2021, pp. 1--5.

\bibitem{Siegel2021}
A.~F. Siegel and M.~R. Wagner, ``Profit estimation error in the newsvendor model under a parametric demand distribution,'' \emph{Manage. Sci.}, vol.~67, pp. 4863--4879, 8 2021.

\bibitem{Jahani2024}
T.~Jahani-Nezhad, P.~Moradi \emph{et~al.}, ``Private, augmentation-robust and task-agnostic data valuation approach for data marketplace,'' \emph{arXiv}, 2024.

\bibitem{Lee2021}
S.~Lee, H.~Kim \emph{et~al.}, ``A data-driven distributionally robust newsvendor model with a wasserstein ambiguity set,'' \emph{J. Oper. Res. Soc.}, vol.~72, pp. 1879--1897, 8 2021.

\bibitem{Liu2020}
J.~Liu, ``Absolute shapley value,'' \emph{ar{X}iv}, 3 2020.

\bibitem{Chhachhi2023}
S.~Chhachhi and F.~Teng, ``On the 1-{W}asserstein distance between location-scale distributions and the effect of differential privacy,'' \emph{arXiv}, 4 2023.

\bibitem{Goncalves2020}
C.~Gon{\c{c}}alves, P.~Pinson \emph{et~al.}, ``{Towards Data Markets in Renewable Energy Forecasting},'' \emph{IEEE Trans. on Sustain. Energy}, p.~1, 2020.

\end{thebibliography}

\end{document}